\documentclass[aps,prd,twocolumn, superscriptaddress]{revtex4}
\usepackage{amssymb, amsmath, graphicx}
\newcommand{\mc}[1]{{\mathcal{#1}}}
\newcommand{\mdl}{\mc{M}}
\newcommand{\mX}{\mc{X}}
\newcommand{\mS}{\mc{S}}
\newcommand{\mD}{\mc{D}}
\newcommand{\Lmax}{\mc{L}_{\rm{max}}}

\newcommand\lsim{\mathrel{\rlap{\lower4pt\hbox{\hskip1pt$\sim$}}
    \raise1pt\hbox{$<$}}}
    
\newcommand{\llmax}[1]{\chi^{2}_{m, (#1)} }    
\newcommand{\avev}{\bar{\Sigma}}    
\newcommand{\pR}{\mc{R}}
\newcommand{\mR}{\mc{R}}
\newcommand{\LL}{\mc{L}}
\newcommand{\Lest}{\widehat{L_\text{max}}}
\newcommand{\Ltr}[1]{\llmax{#1}}    
\newcommand{\like}{{\mc L}}
\newcommand{\lmax}{{\mc L}_\text{max}}
\newcommand{\tmax}{\theta_\text{max}}
\newcommand{\ord}[1]{{\mc O}(#1)}

\begin{document}
\title{Introducing doubt in Bayesian model comparison}

\date{\today} 

\author{Glenn D.~Starkman}\affiliation{CERCA \& Department of
  Physics\\Case Western Reserve University, 10900 Euclid Ave,
  Cleveland, OH 44106, USA}

\author{Roberto Trotta}\affiliation{Astrophysics Group, Imperial
  College London \\ Blackett Laboratory, Prince Consort Road, London
  SW7 2AZ, UK} \affiliation{ Astrophysics Department, Oxford
  University \\ Denys Wilkinson Building, Keble Road, Oxford OX1 3RH,
  UK}

\author{Pascal M.~Vaudrevange}\affiliation{CERCA \& Department of Physics\\Case
 Western Reserve University, 10900 Euclid Ave, Cleveland, OH 44106,
 USA}

\begin{abstract}
There are things we know, things we know we don't know, and then there
are things we don't know we don't know. In this paper we address the
latter two issues in a Bayesian framework, introducing the notion of
doubt to quantify the degree of (dis)belief in a model given
observational data in the absence of explicit alternative models. We
demonstrate how a properly calibrated doubt can lead to model
discovery when the true model is unknown.
\end{abstract}
\maketitle

\section{Introduction}
Given two or more competing models to describe observed data, Bayesian
model comparison offers a way of determining the preferred model given
the data and explicit assumptions about prior beliefs. The key feature
of Bayesian model comparison is that it implements Occam's razor, by
selecting the model that optimally balances quality--of--fit and high
model predictivity.  (See~\cite{Trotta:2008qt} for an introduction.)
Given a set of known models, however, the Bayesian framework usually
has little to say about the absolute quality--of--fit of the preferred
model. This is because the underlying philosophy is that there is
little virtue in rejecting a model if no better alternative is
present.

In the frequentist approach, a popular (absolute) measure of the
goodness--of--fit is given by the $\chi^2$--per--degree--of--freedom
($\chi^2\slash$dof) rule--of--thumb, where $\chi^2$ is (twice) the
best--fit log--likelihood. For normally distributed data points,
$\chi^2\slash$dof is distributed as a $\chi^2$
distribution. Therefore, if the model is appropriate for the data, one
expects that $\chi^2\slash$dof$\approx1$. An unsatisfactory fit is
signaled by $\chi^2\slash$dof$ \gg1$, while $\chi^2\slash$dof$ \ll1$
usually implies overfitting, hence a model overspecification.
Complementary to this, principle component analysis (PCA) can be used
to determine the maximal number of parameters a given observational
data set can reasonably constrain \cite{Huterer:2002hy}. Diagonalizing
the covariance matrix of the parameters and determining how many
eigenvalues are below a given threshold gives an upper limit on the
number of parameters that are well--constrained by observations,
preventing the use of a too--general model. The Bayesian framework
replaces this with the notion of model complexity,
see~\cite{Kunz:2006mc}.

In the Bayesian framework, the question of whether the preferred model
describes the observations ``well enough'' can be phrased as follows:
what is the degree of belief that there are no other ``reasonable''
models that would better describe the observations?  Historically, the
need for fundamentally new physics has often been driven by a poor fit
between data and existing models, at a point where an explicit
alternative was not available.  For example, the development of
General Relativity was driven in part by the need to explain a single
number -- the anomalous perihelion precession rate of Mercury.  The
increasing complexity of data makes it harder to simply evaluate
discrepancies between theory and experiment and decide on their
significance. In light of the increasing usage of Bayesian statistical
techniques, like Markov Chain Monte Carlo (MCMC) algorithms, it would
be advantageous to develop a reliable measure of confidence in the
best--fit model that can deal with today's large data sets and
multi--dimensional parameter spaces. This is particularly true in the
cosmological context, which faces unique difficulties -- some
observations are now so advanced as to be constrained by fundamental
limitations on the quality of data (cosmic variance). Thus,
cosmologists must take particular care to extract the maximum amount
of information from available measurements.

Our confidence in the (absolute) adequacy of the best model can only
be determined under general assumptions about any hypothetical better
fitting model. In this paper, we propose a set of assumptions for such
a model, define the notion of statistical doubt and illustrate its use
by computing the doubt for a toy linear model.  In a future work, we
will apply this tool to evaluate the current concordance model of
cosmology.

First, we give a short review of Bayesian model selection.  We
introduce the notion of doubt, then discuss a technique for the
calibration of the level of false doubt and demonstrate the usefulness
of doubt for model discovery in an application to a toy linear
model. Finally, we present our conclusions.

\section{Bayesian Model Selection}\label{sec:review_ms}

In this section, we briefly review Bayesian model selection. For more
details we refer the reader to \cite{Trotta:2008qt}. From Bayes'
theorem, the posterior probability of model $\mdl_j$ given the data
$d$, $p(\mc{M}_j|d)$, is related to the Bayesian evidence (or model
likelihood) $p(d|\mc{M}_j)$ by
\begin{eqnarray} \label{eq:postM}
 p(\mc{M}_j|d)&=&\frac{p(d|\mc{M}_j)p(\mc{M}_j)}{p(d)}\, ,
\end{eqnarray}
where $p(\mc{M}_j)$ is the prior belief in model $\mc{M}_j$. Here and
in the following, ``model'' denotes a choice of theory, with
specification of its free parameters, $\theta_j$, {\em and} of their
prior probability distribution, $p(\theta_j|\mdl_j)$. The
specification of the prior might be somewhat ambiguous for models with
continuous free parameters, especially when one is working with an
effective parameterisation only loosely tied to the underlying
physics.  (For further discussion of these points,
see~\cite{Liddle:2007ez} and, for a critical
view,~\cite{Efstathiou:2008ed}). In Eq.~\eqref{eq:postM}, $p(d)=\sum_i
p(d|\mathcal{M}_i)p(\mc{M}_i)$ is a normalisation constant (where the
sum runs over all available known models $\mdl_i$, $i=1,\dots, N$) and
\begin{equation} \label{eq:Bayesian_evidence}
  p(d|\mathcal{M}_j)=\int\!d\theta\, p(d|\theta_j, \mc{M}_j) p(\theta_j | \mdl_j)
\end{equation}
is the Bayesian evidence, where $p(d|\theta_j, \mc{M}_j)$ is the likelihood.

Given two competing models, $\mc{M}_0$ and $\mc{M}_1$, the Bayes
factor $B_{01}$ is the ratio of the models' evidences
\begin{eqnarray}
 B_{01}&\equiv&\frac{p(d|\mc{M}_0)}{p(d|\mc{M}_1)}\, ,
\end{eqnarray}
where large values of $B_{01}$ denote a preference for $\mc{M}_0$, and
small values of $B_{01}$ denote a preference for $\mc{M}_1$. The
``Jeffreys' scale'' (Table~\ref{Tab:Jeff}) gives an empirical
prescription for translating the values of $B_{01}$ into strengths of
belief.
\begin{table}[b]
 {\begin{tabular}{l l  l} \hline 
  $|\ln B_{01}|$ & Odds  & Strength of evidence \\\hline 
 $<1.0$ & $\lsim 3:1$ &  Inconclusive \\
 $1.0$ & $\sim 3:1$ &  Weak evidence \\
 $2.5$ & $\sim 12:1$ & Moderate evidence \\
 $5.0$ & $\sim 150:1$ &  Strong evidence \\
\hline
\end{tabular}}
\caption{Empirical scale for evaluating the strength of evidence when
  comparing two models, $\mdl_0$ versus $\mdl_1$ (so--called
  ``Jeffreys' scale'', here slightly modified following the
  prescriptions given in \cite{Gordon:2007xm,Trotta:2008qt}). The
  right--most column gives our convention for denoting the different
  levels of evidence above these thresholds.\label{Tab:Jeff} }
\end{table}

Given two or more models, specified in terms of their parameterisation
{\em and} priors on the parameters, it is straightforward (although
sometimes computationally challenging) to compute the Bayes
factor. Depending on the problem at hand,
semi--analytical~\cite{Trotta:2005ar,Heavens:2007ka} and
numerical~\cite{Mukherjee:2005wg,Feroz:2007kg,Feroz:2008xx} techniques
are available. In the usual case where the prior of the models is
taken to be non--committal (i.e., $p(\mdl_j) = 1/N$), the model with
the largest Bayes factor ought to be preferred. Thus the computation
of $B_{01}$ allows to select one (or a few) promising model(s) from a
set of known models. However, it contains no information about whether
the selected model is actually a good explanation for the data. This
information is contained in $p(\mdl_j|d)$. From Eq.~\eqref{eq:postM},
it is clear that a correct computation of $p(\mdl_j|d)$ requires the
denominator $p(d)$ to be computed from a reasonably complete sum of
models.

We now turn to the question of how to evaluate our absolute degree of
belief in the adequacy of a set of known models.

\section{Bayesian Doubt}\label{sec:doubt}

\subsection{Introducing doubt} 

In light of the observations in the previous section, we seek to
capitalise on the information in $p(\mdl_j|d)$. We introduce the concept
of {\it doubt} $\mD$ to describe in a quantitative way our degree of
(dis)belief in the ability of any known model in a list $\mc{M}_i$
($i=1,\dots,N$) to describe the data. We begin by expanding our space
of models to include an as--yet unknown model $\mX$, which represents
the possibility that the collection of models presently under
consideration is incomplete and that there might be a ``better'' (in a
Bayesian sense) model that we have not yet identified.  We then define
the doubt $\mD=\mD(\{\mdl_i\}|d)$ as the posterior probability of the
unknown model, $p(\mX|d)$, which from Bayes' theorem is given by
\begin{eqnarray}
  \mD\equiv p(\mX|d)&=&\frac{p(d|\mX)p(\mX)}{p(d)}\nonumber\\
  &=&\frac{1}{1+\sum_{i}\frac{p(d|\mc{M}_i)p(\mc{M}_i)}{p(d|\mX)p(\mX)}}\, ,\, \label{eq:doubt}
\end{eqnarray}
where the sum runs over the known models, $i=1,\dots, N$. The prior
for the unknown model is 
\begin{equation}
  p(\mX) = 1 - \sum_{i=1}^N p(\mdl_i) \,.
\end{equation}

Given some openness about the possibility that our list of known
models is incomplete, and given an estimate of the Bayesian evidence
$p(d|\mX)$ for $\mX$, the doubt expresses the posterior probability
that the list of models $\mc{M}_i$ is missing a model that is a better
description of the available data. If $p(\mX)>0$, then ``sufficiently
poor evidence'' for the known models $\mdl_i$ (i.e., $p(d|\mc{M}_i)
\ll p(d|\mX)$) will instill enough doubt to question the
appropriateness of $\mdl_i$. Obviously, assuming {\it a priori} that
the known models exhaust the model space, i.e. $p(\mX)=0$, would leave
no room for doubt: $\mD=0$ independent of the evidence $p(d| \mX)$.

The crucial step in evaluating the doubt is estimating the evidence
for the unknown model, $p(d| \mX)$.  Clearly this quantity cannot be
computed using Eq.~\eqref{eq:Bayesian_evidence}, as this would require
the unknown model to be fully specified in terms of its parameters and
priors.  If this was possible, then $\mX$ could be included in the
list of $\mdl_i$ and would not be unknown in the first place.

Fortunately, even without an explicitly specified mode, but given the
data $d$, we can produce an informed guess as to what the evidence for
a ``good" model should be.  If the evidences of the models on the
table, $\mdl_i$, are poor compared to this value, then the Bayes
factors in favour of the unknown model
\begin{equation}
  B_{xi}\equiv\frac{p(d|\mX)}{p(d|\mdl_i)} \gg 1\, , 
\end{equation}
and as consequently (see Eq.~\eqref{eq:doubt}) the posterior
probability of doubt will increase.

What we are suggesting is in fact a {\em calibration of the absolute
  value of the evidence}. Bayesian model comparison focuses on the
Bayes factor, which indicates the change of our {\em relative}
confidence in the models in light of the observed data. Since the
Bayes factor is the ratio of the models' evidences, the absolute value
of the evidence itself is usually deemed irrelevant. (This is only
actually strictly true for nested models, where the normalisation of
the evidence drops out of the ratio.)  A shortcoming of ignoring the
absolute value of the evidence is that the model comparison will
always return a preferred model, even in cases when all of the
available models fit the data poorly. The notion of doubt is designed
to remedy this obviously unsatisfactory situation, by introducing a
Bayesian way of dealing with the concept of absolute quality of
fit. This is a familiar concept from the usual frequentist
goodness--of--fit tests, which have the advantage of flagging strong
discrepancies between the model and the observed data. Intuitively, it
is sensible that we should start doubting the adequacy of our model(s)
whenever the observed data are in poor agreement with their
predictions. An appropriately calibrated absolute value of the
evidence can be employed within a Bayesian-style reasoning to
substantiate our intuition that ``something fishy'' must be going on
whenever the data are a poor fit to the best model available.

\subsection{Calibration of the evidence} 

The absolute upper bound on the value of the evidence for the unknown
model is achieved for a model $\mS$ which predicts exactly the data
that have been observed (and which has a prior that goes to zero for
any other observation). Such a model can be dubbed a ``sure--thing
model'', because it is totally deterministic. However, in most
situations of interest, such a model is unrealistic, because it does
not allow for the statistical nature of the measurement process, which
is subject to noise, neither does it accommodate a possible statistical
connection between the observables and the underlying physical model,
which introduces sample variance in certain contexts ({\it e.g.}
cosmic variance in cosmology). Furthermore (as discussed
in~\cite{JaynesBook} for the conceptually simple case of coin
tossing), such models are usually thrown out from the beginning,
simply because there is a large number of them: e.g. for any outcome
of $N$ coin flips, there are $2^N$ different sure--thing models
$\mS_i$ ``predicting'' exactly the data that might have been
observed. Because of their large number, each of the $\mS_i$ should be
penalised by a prior probability $p(\mS_i) \sim 2^{-N}$, which goes
quickly to 0 for even moderate values of $N$. For all those reasons,
calibrating off the absolute maximum value of the evidence is
undesirable. A more realistic calibration is required.

We suggest to calibrate the evidence using the properties of the
likelihood and a default (weak) reference prior.  The first step is to
approximate the evidence for the unknown model, $p(d|\mX)$, via the
Bayesian Information Criterion
(BIC)~\cite{Schwarz:1978,Raftery:1995b,Liddle:2004nh,Liddle:2007fy,Trotta:2008qt},
the derivation of which we sketch below (see e.g.~\cite{Raftery:1995b}
for further details).

Let us denote the likelihood of the unknown model by $\like(\theta)
\equiv p(d|\theta, \mX)$ and the prior by $p(\theta|\mX)$. We begin by
Taylor expanding $g(\theta) = \ln
\left[\like(\theta)p(\theta|\mX)\right]$ around the maximum likelihood
value, $\tmax$. To second order,
\begin{equation}
g(\theta) \approx g(\tmax) - \frac{1}{2}(\theta-\tmax)^t H (\theta-\tmax),
\end{equation}
where $H$ is minus the Hessian matrix
evaluated at the maximum likelihood point,
\begin{equation} 
H_{ab} \equiv  -\frac{\partial^2 g(\theta)}{\partial \theta_a \partial \theta_b}{\Big\vert}_{\theta = \tmax}.
\end{equation} 
Using this approximation in the calculation of the evidence,
Eq.~\eqref{eq:Bayesian_evidence}, we obtain
\begin{align} \label{eq:LaplaceApprox}
\ln p(d|\mX) & = \ln \lmax + \ln p(\tmax|\mX) + \frac{k}{2}\ln (2\pi) \\
& -\frac{1}{2} \ln  \vert H \vert + \ord{n^{-1}}\, ,\nonumber
\end{align}
where $k$ is the number of parameters in the unknown model. For large
samples, we can approximate $H \approx n I$ (to order
$\ord{n^{-1/2}}$), where $I$ is the expected Fisher matrix from a
single observation. We now assume that the (unknown) prior
$p(\theta|\mX)$ is a multivariate Gaussian approximately centred at
$\tmax$ with Fisher matrix $I$. This means that the assumed prior
distribution contains about the same (weak) information as would an
average single observation. Then
\begin{equation}
\ln p(\tmax|\mX)  = - \frac{k}{2}\ln (2\pi)  + \frac{1}{2} \ln  \vert I \vert.
\end{equation}
Plugging this reference prior into \eqref{eq:LaplaceApprox} and with
the above approximation for $H$, terms of order $\ord{1}$ cancel and
we obtain
\begin{equation}\label{eq:BICest}
\ln p(d|\mX) = \ln \lmax - \frac{k}{2}\ln n + \ord{n^{-1/2}}.
\end{equation}

This is the standard expression for the BIC, which we will employ to
estimate the evidence for the unknown model. It requires an estimate
of the best--fit likelihood $\lmax$ and of the number of free
parameters, $k$, for the unknown model $\mX$. Notice that the
likelihood, when normalised over the data space, is a dimensionful
quantity, with dimensions $[\mathrm{data}]^{-n}$. In the following we
will always drop such a prefactor (and the associated factors of
$2\pi$) as it always cancels when considering evidence ratios (for the
same data), therefore $\lmax$ has to be regarded as dimensionless.

In order to compute the evidence for the unknown model from
Eq.~\eqref{eq:BICest}, we need to specify an estimator
$\Lest\equiv{-2\ln\widehat{\lmax}}$ for (minus twice) the best--fit
log--likelihood, $-2\ln \Lmax$, of the unknown model $\mX$. This can
be obtained from the requirement of ``typicality'' of the observed
realization under $\mX$. Assuming that the data are normally
distributed, $-2\ln \Lmax$ follows an approximate
$\chi^2$-distribution with $m=(n-k)$ degrees of freedom. Then the
expectation value of $\Lmax$ (as taken over different realizations of
the data, represented by $\langle \cdot \rangle$) follows from
\begin{equation}
 \langle-2\ln \Lmax\rangle=m\, .  
\end{equation}

Employing $\langle-2\ln \Lmax\rangle$ as an estimator for $-2\ln\Lmax$
would be equivalent to assuming that the unknown model has
$\chi^2\slash$dof$=1$, in agreement with the rule--of--thumb for
goodness--of--fit tests. This however is too harsh a requirement on
the performance of the known models $\mdl_i$. Even if one of the known
models is indeed the correct one, the realized maximum likelihood
value for that model will be smaller than the estimator
(i.e.~$-2\ln\Lmax^\text{obs} < \langle-2\ln\Lmax\rangle$) in about
$50\%$ of realizations of the data (for the median and the mean of the
chi--square distribution are very close). This would lead in many
cases to unjustified doubt of the correct model as a consequence of
harmless statistical fluctuations in the observed data realization.

Therefore, instead of using the expectation value, the value of
$\Lmax$ should be more conservatively estimated so that for example,
$-2\ln\Lmax^\text{obs} < \Lest$ only in $100\alpha \%$ of the data
realizations, where we are free to choose the value of $\alpha$. This
can be achieved by taking $\Lest$ to be the $\alpha$ quantile
$\llmax{\alpha}$ of the chi--square distribution with $m$ dof,
$P_{\chi^2_m}$, $\Lest=\llmax{\alpha}$, defined through
\begin{equation}
  \int_{\llmax{\alpha}}^\infty P_{\chi^2_m}(x) {\rm d}x  = 1-\alpha\, .
\end{equation}
As $ \Ltr{\alpha}$ increases monotonically with $\alpha$, larger
values of $\alpha$ lead to smaller (and hence more conservative)
values for the evidence $p(d|\mX)\propto \exp(-\Ltr{\alpha}/2)$ via
Eq.~\eqref{eq:BICest}. In principle, one is free to choose the value
of $\alpha$, and we calibrate it by demanding the wrongful rejection
rate of correct models to be smaller than a given value $\gamma$ (see
Tables~\ref{tab:alphacalibration} and \ref{tab:alphacalibration_mdl0}
below).

In summary, we suggest to use as an estimator of the unknown model's
evidence,
\begin{equation} \label{eq:pluginestimate} 
 \ln p(d|\mX, k, \alpha )= -\frac{\Ltr{\alpha}}{2}  -\frac{k}{2}\ln n\, ,
\end{equation}
where on the left--hand side we have conditioned explicitly on the
number of parameters $k$ of $\mX$, and on the quantile value $\alpha$. 

In a Bayesian spirit, one could treat $k$ and $\alpha$ as
hyperparameters, by specifying a prior and marginalising over them in
the evidence. Here, we investigate the behaviour of doubt in a
Gaussian linear toy model, where $k$ is fixed at a plausible value and
$\alpha$ is chosen by calibrating its value on the fraction of cases
where doubt wrongfully grows.

\subsection{Change in the amount of doubt: independence of prior doubt}

The final quantity that needs to be specified in order to compute the
posterior doubt is the prior doubt, $p(\mX)$. It seems to us that
values $10^{-5} \lsim p(\mX) \lsim 10^{-1}$ might be plausible in many
cases of interest, but higher or lower values are certainly
possible. The interesting question is in fact whether doubt increases
or decreases in the light of the observed data.  We assume that the
prior probability of the known models is equally split among them,
i.e.
\begin{equation}
  p(\mdl_1) = p(\mdl_2) = \dots = p(\mdl_N) =\frac{1-p(\mX)}{N}\, .
\end{equation}
(This simplifying assumption can easily be relaxed.) This leads to the
following expression for the relative change between the prior and
posterior doubt:
\begin{align} \label{eq:pR} 
  \pR & \equiv\frac{\mD}{p(\mX)}=\frac{p(\mX|d, k,\alpha)}{p(\mX)} \\
  & = \left[p(\mX) + (1-p(\mX))
  \frac{\avev}{e^{-\Ltr{\alpha}/2}n^{-k/2}} \right]^{-1}\, ,\nonumber
\end{align} 
where we have defined the average known models' evidence 
\begin{equation}
  \avev \equiv\frac{1}{N}\sum_{i=1}^N p(d|\mdl_i).  
\end{equation}
The doubt grows if $\pR > 1$, i.e. for
\begin{equation} \label{eq:criterium}
 \frac{\avev}{e^{-\Ltr{\alpha}/2}n^{-k/2}} < 1,
\end{equation} 
{\em independently of the prior probability for doubt} (as long as
this is strictly greater than zero).

Let us consider the asymptotic behaviour of the criterion given by
Eq.~\eqref{eq:criterium} for a large number of data points, both under
the assumption that the true model is present in and the assumption
that it is absent from the list of known models.

If the true model $\mdl_T$ is within the list $\mdl_i$, then by
construction of the BIC 
\begin{equation}
 \lim_{n\rightarrow\infty}p(d|\mdl_T)=e^{-\text{BIC}/2} = e^{-\Ltr{0.5}/2} n^{-k/2}\, ,
\end{equation}
hence to leading order 
\begin{equation} \label{eq:leadingorder}
  \frac{\avev}{e^{-\Ltr{\alpha}/2}n^{-k/2}} \rightarrow \frac{1}{N}\exp(\Delta_\alpha/2)\, .
\end{equation}
Here $\Delta_\alpha \equiv \llmax{\alpha}-\llmax{0.5}$, with
$\llmax{\alpha}$ the inverse of the $\chi^2$ cumulative distribution
function (CDF) with $m$ degrees of freedom. Clearly, $\Delta_\alpha >
0$ for $\alpha>0.5$. In the limit of many data points,
$n\rightarrow\infty$ (or equivalently $m\rightarrow\infty$),
$\llmax{\alpha}$ can be approximated by the inverse of the CDF of the
normal distribution with mean $m$ and variance $2m$, $\mc{N}(m, 2m)$,
so that
\begin{eqnarray} \label{eq:erflimit}
 \lim_{n\rightarrow\infty}\Delta_{\alpha} = m\times \left(2\sqrt{2}\,\text{InverseErf}(2\alpha-1)\right)\, ,
\end{eqnarray}
where $\text{InverseErf}(x)$ is the inverse of the Error Function. It
follows that $\Delta_\alpha \gg 1$ and therefore from
(\ref{eq:leadingorder}) and (\ref{eq:pR}) for many data points $n$ (or
degrees of freedom $m$), $\pR\rightarrow 0$, and hence $\mD
\rightarrow 0$. In other words: if the true model is in the list of
known models, the doubt goes to zero as expected. Notice that in
Eq.~\eqref{eq:leadingorder} the extra factor $1/N$ of penalising for
the true model comes from the fact that its predictivity has been
spread among a set of $N$ possibilities. More precisely, the $1/N$
factor assumes that the evidences for the other known models are
negligible in the sum. However, if there are $M<N$ other models which
are unnecessarily more complicated than the true model with parameters
that are unconstrained by the data, one would expect that the evidence
for each of those models is of the same order as for the true model
(because the evidence does not penalise unconstrained
parameters). Therefore the $1/N$ factor would be replaced by a factor
$\sim (M+1)/N$.

If, instead, the true model (or another model that is about as good as
the true model in explaining the data) is not within the list of known
models, then the numerator in Eq.~\eqref{eq:criterium} will drop very
quickly to zero, hence
\begin{equation}
  \lim_{n\rightarrow\infty}\pR = \frac{1}{p(\mX)}\, .
\end{equation}
Therefore the doubt goes to unity, $\mD\rightarrow 1$, which leads to
questioning the completeness of our list of known models.

Further modelling requires the explicit specification of the known
models and computation of $\avev$ from the observed data. We therefore
proceed with an illustration based on linear models.

\section{Illustration: Linear Toy Model}
\label{sec:toymodel}

It is instructive to look at an example for the usage of doubt in a
simple toy model. Consider the case of a {\em Gaussian linear model}:
\begin{equation}
  y = A \theta + \epsilon\, ,
\end{equation}
where the dependent variable $y$ is a $n$-dimensional vector of
observations, $\theta$ is a vector of dimension $c$ of unknown
regression coefficients and $A$ is a $n\times c$ matrix of known
constants which specify the relation between the input variables
$\theta$ and the dependent variables $y$. Furthermore, $\epsilon$ is a
$n$-dimensional vector of random variables with zero mean (the {\em
  noise}). We assume that the observations are independent identically
distributed (i.i.d.), hence $\epsilon$ follows a multivariate Gaussian
distribution with unit covariance and the likelihood is given by
\begin{equation}
  \LL(\theta)=exp\left[-\frac{1}{2}\sum_{i=1}^n\left(\frac{y_i - y_i^{\text{th}}}{\sigma}\right)^2\right]\, ,
\end{equation}
where $y_i (y_i^{\text{th}})$ are the values of the observed (predicted)
observables, and $\sigma=1$.

For the purpose of our example, let us assume that we have $n$ data
points, and that two models are available:
\begin{itemize}
\item $\mc{M}_0: y = \theta $, i.e. $c=1$ and $A=(1,\dots,1)^t$ and
\item $\mc{M}_1: y = \theta x$, i.e. $c=1$ and $A=(x_1,\dots,x_n)^t$ .
\end{itemize}
In both cases there is one free parameter, $\theta$, and we will
assume that a prior is available of the form $p(\theta|\mdl_0) =
p(\theta|\mdl_1) = 1/4$ for $\theta\in[-2,2]$ (and vanishes outside
that range). We will assume that there is one free parameter in the
unknown model, {\it i.e.} $k=1$. (The number of effective free
parameters can be investigated further by the mean of the Bayesian
complexity, see~\cite{Kunz:2006mc}.)

We are interested in investigating the behaviour of the ratio of the
posterior to the prior doubt $\pR$. We expect $\pR<1$ (decreasing
doubt) when the known model is the correct underlying distribution,
and $\pR>1$ when an incorrect known model is used. For definiteness,
we will take the true model to be $\mdl_1$, with
$\theta_\text{true}=0.1$.
 
\subsection{Doubt when fitting data with the correct model: false doubt}
Let us assume that our list of known models contains only $\mdl_1$
({\it i.e.} $N=1$ and $\mdl_0$ is not on the list). We first fit the
dataset (generated from $\mdl_1$) with the correct model $\mdl_1$ and
we compute the posterior $p(\mdl_1|d)$ and the doubt $p(\mX|d)$ from
Eq.~(\ref{eq:doubt}), using $\alpha=0.95$
\begin{figure}
  \includegraphics[width=\linewidth]{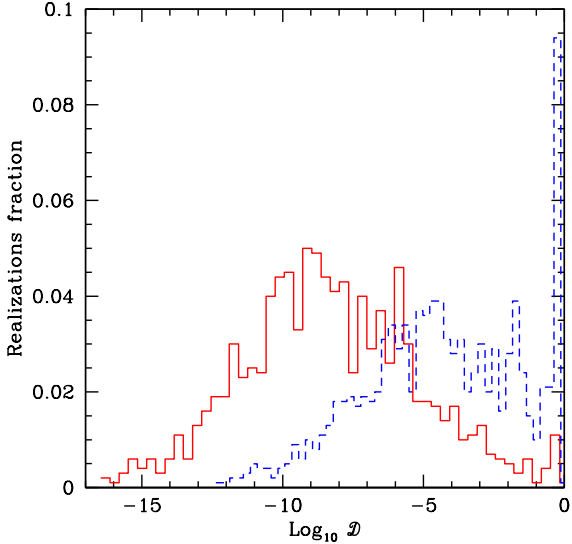}
  \caption{Distribution of doubt values when the correct
    model is used to fit the data, for $1000$ realizations of $100$
    data points, for prior doubt $p(\mX)=10^{-1}$ (blue, dashed) and
    $p(\mX)=10^{-5}$ (red, solid), and $\alpha=0.95$.}
  \label{fig:nodoubt_histo}
\end{figure}
In this case, there should be no reason for doubt as we expect
$\mdl_1$ to be an adequate description of the data. We show the
ensuing distribution of posterior doubt in
Fig.~\ref{fig:nodoubt_histo} (from 1000 data realizations with $n=100$
data points each), for two different choices of prior doubt, $p(\mX) =
10^{-5}$ (red/solid histogram) and $p(\mX) = 10^{-1}$ (blue/dashed
histogram). The posterior doubt of the vast majority of the
realizations is smaller than the prior doubt, consistent with
expectations. Clearly, the absolute value of the posterior doubt
depends on the choice of prior doubt, and quite reasonably so. If {\em
  a priori} one is quite certain that the model is correct, then small
deviations from a perfect fit will not shake one's belief in the
model. However, if the prior doubt is relatively large, $p(\mX) =
10^{-1}$ (i.e., if {\em a priori} one is quite uncertain that the
model being used is correct) then already small fluctuations in the
data will lead to relatively strong posterior doubt. The less
confident one is to begin with, the more easily one's belief in a
model is shaken by statistical fluctuations. In any case, larger
amounts of data will lessen the effect of fluctuations, leading to
little doubt about the correct model, with the amount of data required
for persuasion dependent on the prior doubt. For both values of prior
doubt in Fig.~\ref{fig:nodoubt_histo}, about $14\%$ of the
realizations lead to a posterior doubt that is larger than the prior
doubt. This number is independent of the prior doubt as can be seen
from Eq.~\eqref{eq:pR} and decreases with increasing number of data
points $n$ and as $\alpha$ approaches unity. We call realizations that
incorrectly give an increase of doubt (although the model being used
is the correct one) cases of ``false doubt''.

\begin{figure}
  \includegraphics[width=0.95\linewidth]{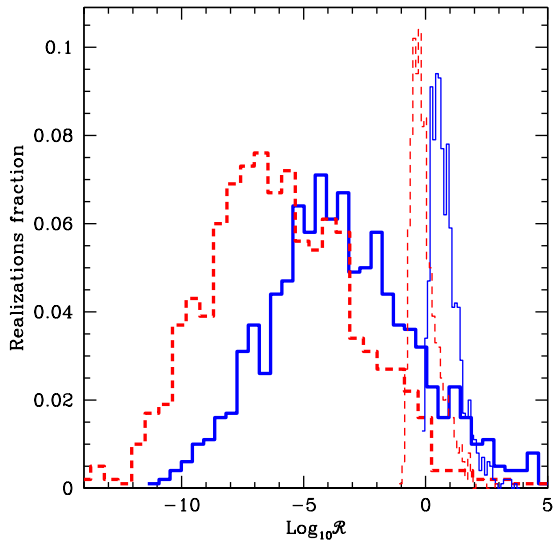}
  \caption{Distribution of the change in doubt $\log\mR$ when fitting
    data with the correct model, from 1000 data realizations for
    $n=5$, $\alpha=0.95$ (blue/solid, thin), $n=5,\alpha=0.99$
    (red/dashed, thin), $n=100,\alpha=0.95$ (blue/solid, thick) and
    $n=100,\alpha=0.99$ (red/dashed, thick).  Values $\log\mR < 0$
    correspond to a decrease in the amount of doubt. A larger value of
    $\alpha$ and a larger number of data points $n$ lead to a
    reduction of the number of cases where the doubt wrongly grows
    (values $\log\mR>0$).  }
  \label{fig:RTlinearMlinearCombined}
\end{figure}

We now turn to investigate the relative change in doubt, $\mR$, which
is plotted in Fig.~\ref{fig:RTlinearMlinearCombined}, for two choices
of the number of data points $n$ and of the calibration parameter
$\alpha$. Recall that $\log \mR > 0$ ($<0$) corresponds to an increase
(decrease) in doubt in light of the observed data. In fact, $\mR$ can
be regarded as a sort of ``Bayes factor'' for doubt change --- it
gives the relative change in our ``state of doubt'' after we have seen
the data.  While the actual value of $\mR$ is dependent on the prior
doubt, the threshold separating increasing doubt from decreasing doubt
(i.e., $\log \mR = 0$) is independent of $p(\mX)$. As expected, a
larger number of data points leads to a decrease in the fraction of
realizations for which doubt wrongfully grows (values
$\log\mR>0$). The same is true if one increases $\alpha$.  As
explained above, this is because a larger value of $\alpha$ leads to a
less harsh penalty for odd features in the realized data.

\begin{figure}
  \includegraphics[width=\linewidth]{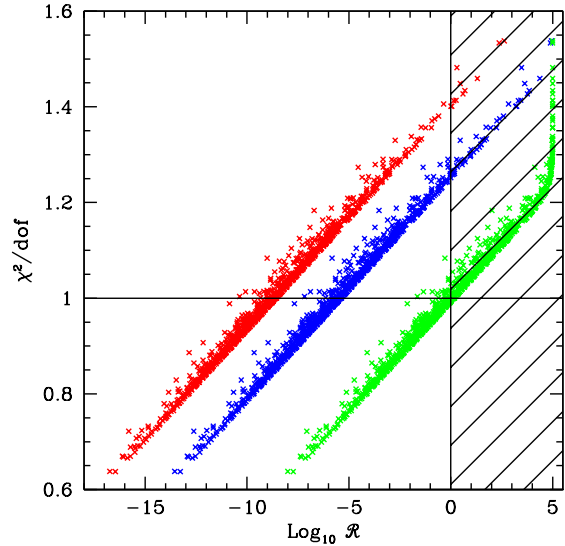}
  \caption{Correlation between the ``chi--square--per--dof'' rule and
    the change in doubt, $\log\mR$, for different values of the
    calibration parameter $\alpha$, increasing from right to left (for
    1000 data realizations). The parameter $\alpha$ can be chosen so
    that only a pre--determined fraction $\gamma$ of data realizations
    lie in the ``false doubt'' zone (shaded, $\log\mR>0$). Here,
    $\alpha$ has been chosen in such a way that (from right to left)
    $\gamma = 0.50, 0.05, 0.01$.}
  \label{fig:scatterplot}
\end{figure}

By construction of the doubt, there is a strong correlation between
the value of $\log\mR$ and the chi--square per dof of the best
fit. This is depicted in Fig.~\ref{fig:scatterplot}. First, it is
obvious that for a given choice of $\alpha$, data realizations leading
to a wrongful increase in doubt ($\log\mR > 0$) are the ones that
present ``unlucky'' features, i.e. the ones with a large value of
$\chi^2/\text{dof}$. In other words, cases of false doubt would be
suspicious even using a more traditional measure of the quality of
fit. However, the second, crucial point is that the parameter $\alpha$
can be calibrated in order to achieve a pre--determined fraction of
false doubt from a known model. By increasing $\alpha$, the locus of
the realizations shifts to the left of the plot. Therefore one can
choose $\alpha$ in such a way that the probability of false doubt is
below a given threshold. This is discussed in the next section.

\subsection{Calibration of the level of false doubt}

As shown above, some fraction of data realizations will always lead to
false doubt. This fraction depends on the value of $\alpha$ and on the
number of data points, $n$, but not on the level of prior
doubt. (There is a further, if subdominant, dependence on $k$.)

For a given number of data points, it is desirable to tune the value
of $\alpha$ such that the fraction of false doubt $\gamma$ is (on
average) below a predetermined threshold. This is achieved as
follows. Starting from the model distribution (here, $\mdl_1$), we
employ current data to derive constraints on its free parameters, as
usual in the inference step. We then select an estimator
$\hat{\theta}$ for the value of the parameters (here, $\theta$), which
will usually be either the best--fit point or the posterior mean. We
simulate $10^4$ realizations of the data from the model, assuming a
fiducial value $\hat{\theta}$ for its parameters. We then compute the
doubt for each realization, and calibrate the value of $\alpha$ by
requiring that the fraction of realizations with $\log\mR>0$ be below
a value $\gamma$~\footnote{Strictly speaking, this procedure
  underestimates the uncertainty in the spread of values for the
  doubt, in that it ignores the current posterior uncertainty about
  $\hat{\theta}$. Such an approximation will be valid whenever the
  scale of the posterior width is much smaller than the width of the
  prior on the model's parameters (i.e., for informative data). A
  fuller calculation accounting for this extra layer of complexity
  will be presented elsewhere.}. To further reduce the scatter in
$\alpha$, we average over the resulting $\alpha$ values from $1000$
such procedures. Table~\ref{tab:alphacalibration} shows such values of
$\alpha(\gamma, n,\mdl_1)$ for a few representative choices of
$\gamma$ and $n$. One striking feature of the calibration table is
that the value of $\alpha$ required for a false doubt rate $\gamma$ is
systematically much larger than $1-\gamma$. This is another reflection
of the well known fact that how likely the data are given the
hypothesis does not by itself determine how probable the hypothesis is
given the data. Inferring the latter requires the use of Bayes
theorem. (For an in--depth discussion of this point,
see~\cite{Gordon:2007xm}).

\begin{table}
\begin{center}
\begin{tabular}{ l | l l l }
 $n$ & $\gamma = 0.01$  & $\gamma = 0.05$ & $\gamma =0.50$ \\\hline
 10     &   0.99969(3)  & 0.9980(1) & 0.9550(8)\\
 100   &   0.9980(2)  & 0.9869(7) & 0.747(4)\\
 200   &  0.9968(3)   & 0.9810(9) & 0.684(4)\\
1000  &  0.9942(6)   & 0.968(1)   & 0.586(5)
\end{tabular}
\end{center}
\caption{Values of $\alpha(\gamma, n,\mdl_1)$, as a function of the
  number of data points, $n$, ensuring an average fraction of false
  doubt of $\gamma=1\%, 5\%, 50\%$ and for model distribution
  $\mdl_1$. The number in brackets denotes the uncertainty in the last
  digit.}
\label{tab:alphacalibration}
\end{table}

\begin{figure}
 \includegraphics[width=0.95\linewidth]{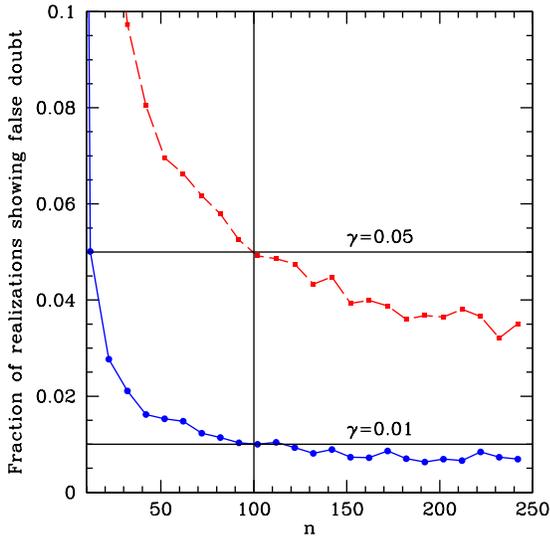}
  \caption{Fraction of cases of false doubt, $\log\mR>0$ as a function
    of the number of data points, $n$, employing a doubt calibration
    parameter $\alpha$ corresponding to a false doubt probability of
    $\gamma = 5\%$ (dashed red) and $\gamma=1\%$ (solid blue) for
    $n=100$. For a number of data points $n>100$ there is a residual
    (if mild) $n$ dependence in the fraction of false doubt, which
    however is always below the calibration level. The calibration is
    independent of the prior doubt.}
  \label{fig:pDoubt}
\end{figure}

Because the calibrated value of $\alpha$ decreases monotonically with
increasing $n$, the above calibration procedure, once carried out for
a certain number of data points, is expected to be conservative when
the amount of data increases. This is shown in Fig.~\ref{fig:pDoubt},
where we plot the fraction of realizations leading to false doubt
after $\alpha$ has been calibrated at $n=100$. We can see that for
$n>100$ the fraction of false doubt remains below the calibrated
level, and that the residual $n$ dependency is fairly mild. 

\subsection{Doubt when fitting data with an incorrect model: model discovery}

We now pretend that the true model where the data come from, $\mdl_1$,
is unknown to us. We take $\mdl_0$ to be the only known model, and
consequently fit the data with it. We repeat the calibration procedure
for $\alpha$ for the known model $\mdl_0$. The corresponding
calibrated values of $\alpha(\gamma, n,\mdl_0)$ are given in
Table~\ref{tab:alphacalibration_mdl0}.

\begin{table}
\begin{center}
\begin{tabular}{ l | l l l }
 $n$ & $\gamma = 0.01$  & $\gamma = 0.05$ & $\gamma =0.50$ \\\hline
2  & 0.9944(5) & 0.973(1) & 0.781(2) \\
3  & 0.9938(6) & 0.969(1) & 0.692(3) \\ 
4  & 0.9935(6) & 0.967(2) & 0.654(4) \\ 
5  & 0.9933(7) & 0.966(2) & 0.633(4) \\ 
6  & 0.9932(7) & 0.965(2) & 0.618(4) \\ 
7  & 0.9931(7) & 0.965(2) & 0.607(4) \\ 
8  & 0.9929(7) & 0.964(2) & 0.600(4) \\ 
9  & 0.9928(7) & 0.963(2) & 0.593(4) \\ 
10 & 0.9928(7) & 0.963(2) & 0.588(5)
\end{tabular}
\end{center}
\caption{Values of $\alpha(\gamma, n,\mdl_0)$, as a function of the
  number of data points, $n$, ensuring an average fraction of false
  doubt of $\gamma=1\%, 5\%, 50\%$ and for the known model
  $\mdl_0$. The number in brackets denotes the uncertainty in the last digit.}
 \label{tab:alphacalibration_mdl0}
\end{table}

\begin{figure}
  \includegraphics[width=\linewidth]{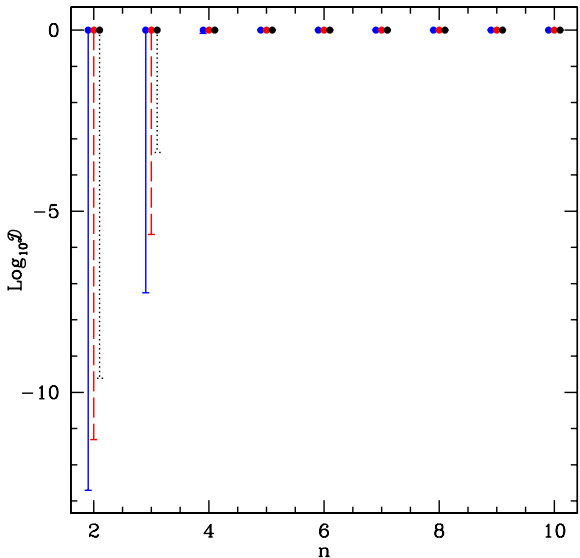}
  \caption{Model discovery: posterior doubt on the (wrong) model
    $\mdl_0$ as a function of the number of data points, $n$, here for
    prior doubt $p(\mX) = 10^{-5}$ using the value of $\alpha$
    calibrated to a fraction of false doubt $\gamma=1\%$ (solid blue
    line), $\gamma=5\%$ (dashed red line), $\gamma=50\%$ (dotted black
    line). The points give the mean doubt over $1000$ data
    realizations and the vertical bars indicate the range of values
    enclosing 95\% of the realizations. The posterior doubt goes very
    quickly to 1 (in fact, for $n \geq 5 $, all realizations have a
    posterior doubt of unity), therefore leading to doubt the
    correctness of the model.}
  \label{fig:discovery_calibrated}
\end{figure}

Using the calibrated values of $\alpha$ we then compute the posterior
doubt on $\mdl_0$. The result is shown in
Fig.~\ref{fig:discovery_calibrated} (again averaged over $1000$
realizations). We can see that doubt increases immediately from its
prior value $p(\mX) = 10^{-5}$, and tends very quickly to $1$. This
clearly signals the inadequacy of the known model to fit the data. We
should therefor question the correctness of the known model and
suspect the existence of a better model. Therefore our procedure leads
to model discovery in the absence of an explicit specification of the
alternative, true model.

\subsection{Generalisation to multi-models cases and discussion} 

In the example considered so far, we have only dealt with doubt (or
its absence) for one model at a time.  The situation is qualitatively
similar when several alternative known models are available (i.e. for
$N>1$).

When several known models exist, the calibration procedure should be
carried out on the model that is currently the best among them,
i.e. on the model with the largest Bayes factor. This ensures that the
probability of false doubt is under control for the currently favoured
model (which has the largest evidence). If one of the known models is
clearly preferred, then the situation is qualitatively similar to the
case of $N=1$ (since the evidence from the other, poorer models
contributes very little to the sum in the definition of doubt). If
instead several models have a similar value of the evidence under
present data, then it is expected that the outcome of the calibration
should be quite similar for any of them. Also, as discussed above in
the presence of $M$ models with approximately the same evidence there
is an extra ``volume factor'' $M/N$ to take into account in the rate of
false doubt.

The false doubt calibration procedure introduced here insures against
unjustified doubt of the known models at a given threshold (set by
$\gamma$). Because unknown models belong to the world of unknown
unknowns, it is more difficult to calibrate the performance of doubt
for model discovery, {\it i.e.} for justifiably doubting false models.
Whether or not the doubt does increase when the true model is
genuinely unknown depends on how different that unknown true model is
from the known models. Here, ``different'' must be interpreted in
terms of a ``distance'' in the space of models, as measured by the
Bayesian evidence. In this sense, the notion of doubt introduces an
absolute metric in model space, to complement the relative metric
represented by the Bayes factor. In any case, if the true model is not
very different in its observational consequences from one of the known
models (in which case doubt will not increase), then one might
conclude that the known model is a phenomenologically accurate
description of the presently available data. If doubt does increase,
though, this is a signal that the best available model is an
inadequate description of the observations and that new theoretical
input is required.

In general, we remark that the existence of known models that are
unnecessarily complex (i.e., with more free parameters than the true
model) is not directly addressed by doubt. In this case, an analysis
of doubt (properly calibrated) will return a null result, i.e. no
reason for doubting the adequacy of the overly complex model. One has
to keep in mind that doubt is a tool for {\em model discovery}, whose
primarily goal is to point towards the need to enlarge (or change
completely) the space of the known models. The shedding of unnecessary
levels of complexity is instead a task best accomplished by
simultaneously analysing the evidence and Bayesian complexity.
(See~\cite{Kunz:2006mc} for an application.)

Finally, the usual caveats apply about the dependence on the volume in
parameter space enclosed by the parameters' prior $p(\theta_j |
\mdl_j)$, as is always the case for calculations involving the
Bayesian evidence.  (See~\cite{Trotta:2006ww,Trotta:2008qt} for a
discussion.) However, the calibration procedure for $\alpha$
automatically accounts for the volume enclosed by the chosen prior
under $\mdl_j$, as compared with the reference prior employed for the
unknown model. If one were to change the prior on the parameters of
the known model, then this would amount effectively to a change of
model.  (As mentioned, we consider a model specification to consist of
both the model parameters {\em and} their prior.) Therefore the
calibration ought to be performed again on the new model.

\section{Conclusions}\label{sec:conclusions}

Checking the appropriateness of a given set of models to describe
observations is not a usual task in a Bayesian framework.  We have
suggested an intuitive approach to doubt in a Bayesian context that
shares some philosophy with the frequentist approach: after all, the
estimator for $\Lmax$ is based on a $\chi^2\slash$dof argument and
strictly speaking, Bayesians should show little interest in the
hypothetical outcome of different realizations of
reality. Nevertheless, as we demonstrated with the example of a simple
linear model, the concept of doubt is more powerful than traditional
goodness--of--fit tests provided the parameter $\alpha$ controlling the
rate of false doubt is correctly calibrated.

As mentioned in the introduction, the concept of doubt is ideally
suited for applications in cosmology. Huge data sets and
multi-dimensional parameter spaces do not lend themselves very well to
visual inspection. Computing the doubt $\mD$ -- a single number --
gives an indication of the trustworthiness of the model(s) under
consideration in a Bayesian context. Applications range from questions
about the very early inflationary phase of the universe (in particular
about the shape of the primordial power spectrum generated during
inflation) to the future evolution of the Universe which appears to be
dominated by dark energy (in particular whether the equation of state
during late--time acceleration is constant). Given the need to
calibrate the doubt, such work requires a large amount of
computational power even given recent advances in numerical techniques
for the evaluation of the evidence, and it will be addressed in a
future paper.

\section*{Acknowledgements}
We would like to thank Pietro Berkes, Francesc Ferrer, Andrew Jaffe,
Irit Maor and John Ruhl for stimulating discussions. R.T. was
partially supported by the Royal Astronomical Society through the Sir
Norman Lockyer Fellowship, by the Science and Technology Facilities
Council (UK) and by St Anne's College, Oxford. R.T. would like to
thank CERCA for the hospitality in the early phases of this
work. G.D.S and P.M.V. are supported by a grant from the DOE to the
particle astrophysics group at CWRU. P.M.V. is supported by the office
of the Dean of the College of Arts and Science at CWRU.

\bibliographystyle{h-physrev}

\bibliography{doubt}

\end{document}